\documentclass[11pt,a4paper]{article}
\usepackage[utf8]{inputenc}
\usepackage[T1]{fontenc}
\usepackage{lmodern}
\usepackage{microtype}

\usepackage{amsmath, amssymb, amsthm}
\usepackage{graphicx}
\usepackage{booktabs}
\usepackage{hyperref}
\usepackage{cleveref}
\usepackage{url}
\usepackage{geometry}
\usepackage{xcolor}
\usepackage{algorithm}
\usepackage{algpseudocode}
\usepackage{subcaption}

\theoremstyle{definition}

\hypersetup{
    colorlinks = true,
    linkcolor  = blue!70!black,
    citecolor  = green!50!black,
    urlcolor   = cyan!70!black
}

\title{\textbf{Constructing Inverse Potentials from Scattering Phase Shifts using Physics-Informed Neural Networks: Application to Neutron-Alpha Scattering}}

\author{
  Ayushi Awasthi$^{1}$,\;
  Ishwar Kant$^{1}$,\;
  Arushi Sharma$^{1}$,\;
  M.~R.~Ganesh Kumar$^{2}$,\;  \\ and
  O.~S.~K.~S.~Sastri$^{1,*}$
  \\[8pt]
  \small
  $^{1}$Department of Physics and Astronomical Sciences,\\
  \small Central University of Himachal Pradesh, Dharamshala, India\\[4pt]
  \small $^{2}$Applied Materials India Private Limited,
         Bengaluru 560066, India\\[6pt]
  \small $^{*}$Corresponding author : sastri.osks@hpcu.ac.in
}
\date{}   

\begin{document}

\maketitle
\begin{abstract}
We develop a physics-informed neural networks (PINNs) framework for the
inverse scattering problem in nuclear physics and apply it to the
$P_{3/2}$ partial wave of neutron-alpha elastic scattering.
The radial potential is represented by a feed-forward network whose
output is multiplied by a Gaussian envelope, embedding the finite-range
condition directly into the architecture rather than through a soft
penalty term. This distinction proves essential: without the envelope, the optimizer produces potentials with non-vanishing tails and the resulting phase shifts remain inconsistent with the data regardless of training
duration, demonstrating that hard structural constraints are
indispensable for physically meaningful solutions to nuclear inverse
problems. Phase shifts are generated at each scattering energy by numerically
integrating the variable-phase equation with a fourth-order Runge-Kutta
scheme, making the entire pipeline end-to-end differentiable.
Training converges stably to a loss near $3\times10^{-4}$ and recovers
a smooth, purely attractive central potential with a well depth of
 $-60.47$~MeV. Adding the centrifugal barrier to the learned potential reveals a well-defined barrier-well structure that naturally accounts for the $P_{3/2}$ resonance. The extracted resonance parameters, $E_{r} = 0.95$~MeV and $\Gamma_{r} = 0.78$~MeV, together with the P-wave effective-range parameters, are in good agreement with expected values. A leave-one-out analysis confirms that the reconstruction is stable
against the removal of any single data point. These results establish physics-guided machine learning as a reliable route to potential reconstruction from nuclear scattering data.
\end{abstract}
\section{\label{sec:level1}Introduction}
Nuclear scattering experiments have long served as one of the most important experimental tools for probing the nature of nuclear forces \cite{arrington2022progress}. When two nuclei collide at low to intermediate energies, the angular distribution and energy dependence of the scattered particles carry detailed information about the underlying interaction potential \cite{Newton1982}. Extracting that potential from measured data, the so-called inverse
scattering problem, has therefore been a central goal of nuclear
physics since the earliest days of the field~\cite{Newton1982,Chadan1977}.
The importance of this problem extends well beyond pure theory: reliable
nuclear potentials are essential inputs to reaction network calculations
in astrophysics, serve as benchmarks for ab initio many-body
methods, and find direct application in nuclear data
evaluation~\cite{Machleidt2011}. Despite this long history, the inverse problem remains genuinely difficult. It is inherently ill-posed, many distinct potentials can reproduce the same finite set of phase shifts, and any method that aspires to be reliable must impose additional physical information to render the solution unique and stable.
Among the lightest nuclear systems, the neutron–alpha (n–$\alpha$) scattering process is one of the simplest nucleon–nucleus systems, as it involves a neutral projectile and a tightly bound, spin-zero $\alpha$ particle, whose internal degrees of freedom are effectively frozen at low scattering energies, reducing the problem to a single nucleon interacting with an inert core.
The phase-shift data for this system have been available since the late 1960s over a broad energy range, with the measurements of Satchler~et~al.\ providing a benchmark dataset spanning laboratory energies from well below 1~MeV to approximately 18~MeV~\cite{Satchler1968}. Despite its simplicity, the system exhibits nontrivial dynamics, including the well-known $^{5}$He resonance in the $P_{3/2}$ channel near $E_r \approx 0.92$ MeV (c.m.), which provides a sensitive test of the P-wave interaction~\cite{Tilley2002,Nollett2007}.
Several ab initio calculations, including recent chiral
effective field theory approaches, have attempted to describe the
n-$\alpha$ system from first principles~\cite{Lynn2016,Nollett2007},
but constructing a simple, physically transparent effective potential
directly from scattering data remains a relevant and instructive problem
in its own right. Efforts to reconstruct the n-$\alpha$ interaction from phase shifts have a long history, and several methods have been developed with varying degrees of success.
Phenomenological optical models represent the potential as a parametric
function, most commonly in the Woods-Saxon form~\cite{Hodgson1994},
with parameters adjusted until the computed phase shifts or cross
sections match experimental data. This approach is straightforward and widely used, but the choice of functional form introduces a bias that is difficult to quantify: if the true potential does not resemble the assumed well shape, the fit may remain accurate only within the energy range used for fitting and fail to predict properties outside it.

R-matrix theory~\cite{Wigner1947,Descouvemont2010} offers a more
systematic framework by expanding the interior wave function in terms
of energy-independent basis states defined within a finite matching
radius. It connects naturally to reaction theory and has been applied
extensively to light nuclear systems, but still requires choosing the
channel radius and the number of basis states, introducing model
dependence at another level.
For a system like $^{5}$He, where the resonance is broad and overlaps
with the continuum, convergence of the pole expansion is not guaranteed
across the full energy range of interest.
Direct inversion methods, such as those based on the Gel'fand-Levitan
or Marchenko integral equations~\cite{Chadan1977}, are exact in
principle: given complete phase-shift information over all energies,
they uniquely determine the potential.
In practice, however, experimental data are available only at discrete
energies and over a finite energy range, so the input must be
supplemented by theoretical assumptions at unmeasured energies.
The resulting potential is sensitive to these assumptions, and the
propagation of experimental uncertainties through the inversion kernel
is technically demanding.
Recently our group has addressed the reconstruction problem for several
nuclear scattering systems using a genetic algorithm combined with a reference potential approach, recovering inverse potentials for $\alpha$-$^{12}$C, nucleon-deuteron, $n$-$p$, $p$-$p$, and $\alpha$-deuteron
scattering~\cite{awasthi2024high,awasthi2025genetic,awasthi2025genetic1,sharma2025novel,sharma2026genetic}. That body of work demonstrates that
population-based optimization can identify physically meaningful potentials
even for resonant and near-threshold systems. However, metaheuristic
methods of this kind are computationally intensive and, crucially, they lack
analytical gradients. The absence of gradient information limits scalability
to higher-dimensional parameter spaces and makes systematic uncertainty
quantification through standard statistical tools considerably harder.
These limitations motivate a gradient-based alternative. Over the past
decade, machine learning has permeated nuclear and hadronic
physics~\cite{Boehnlein2022}, finding applications that range from
interpolating nuclear masses~\cite{Neufcourt2019} and emulating density
functionals~\cite{Utama2016} to accelerating coupled-cluster and no-core
shell-model calculations~\cite{Negoita2019}. In nuclear reactions
specifically, machine learning emulators have been constructed for optical
model cross sections~\cite{Lovell2017} and astrophysical
$S$-factors~\cite{Negoita2018}, and neural network wave functions have
recently been used to solve the many-body Schr\"odinger equation for light
nuclei~\cite{Adams2021}. Within this broader trend, physics-informed neural
networks (PINNs)~\cite{Raissi2019,Lagaris1998} have emerged as a
particularly compelling tool for inverse problems governed by differential
equations. The core idea is to embed the governing equation directly into
the training objective, so the network simultaneously fits the data
\emph{and} learns a solution that is consistent with the known physics. A
comprehensive review of these methods is given by Cuomo
\textit{et al.}~\cite{Cuomo2022}.

A question that has received almost no systematic attention is \emph{how}
physical knowledge should be encoded in the network architecture. Most PINN
implementations express physical constraints as soft penalty terms in the
loss function, which merely discourage rather than prevent constraint
violations. For nuclear scattering, however, the finite-range condition is
not an optional convenience but a fundamental requirement of the scattering
formalism itself. Encoding it only as a penalty leaves open the possibility
of solutions that achieve a low loss value while exhibiting qualitatively
wrong long-range behavior, a failure mode we document explicitly in this
work.
\\ In this work, we have developed a Physics-Informed Neural Network (PINN) framework for the reconstruction of inverse potentials in nucleon-nucleus scattering, and applied it to the $P_{3/2}$ partial wave of neutron-alpha ($n-\alpha$) scattering. Upon obtaining the learned central potential, we construct the effective potential by incorporating the centrifugal term. The partial cross section is subsequently computed, from which the resonance energy and width are extracted. Furthermore, the low-energy P-wave effective-range parameters are calculated. To assess the robustness of the reconstructed potential, a comprehensive sensitivity analysis is performed using a leave-one-out cross-validation procedure. All extracted quantities are compared with available experimental data and established theoretical predictions.

\section{Methodology}
\label{sec:theory}
We describe the n-$\alpha$ elastic interaction by an effective local central potential $V(r)$ in a single partial wave with orbital angular momentum $\ell$. Since the $^{5}$He ground state is dominated by the $P_{3/2}$ resonance, we restrict the present analysis to the $\ell = 1$ channel and treat the potential as purely central; spin-orbit coupling, which would further split the $P_{1/2}$ and $P_{3/2}$ phase shifts, is beyond the scope of this work and is left for a future extension. To compute the scattering phase shifts from $V(r)$ we use the variable-phase approach of Calogero~\cite{Calogero1967,Calogero1963}, which recasts the radial Schr\"{o}dinger equation as a first-order nonlinear initial-value problem. This formulation has a number of practical advantages over direct solution of the Schr\"{o}dinger equation followed by asymptotic matching, and it is particularly well adapted to the present inverse problem, as we discuss below.

For a given scattering energy $E$ and partial wave $\ell$, the variable-phase approach defines a radial phase function $\delta(r;E)$ satisfying \cite{palov2021vpa}
\begin{equation}
  \frac{d\delta(r;E)}{dr}
  = -\frac{1}{k}\,V(r)
    \Bigl[\cos\delta(r;E)\;\hat{j}_{\ell}(kr)
         -\sin\delta(r;E)\;\hat{n}_{\ell}(kr)\Bigr]^{2},
  \label{eq:phase_eq}
\end{equation}
with the boundary condition $\delta(0;E)=0$.
Here
\begin{equation}
  k = \frac{\sqrt{2\mu E}}{\hbar}
\end{equation}
is the relative wave number in the centre-of-mass (CM) frame,
$\mu = m_{n}m_{\alpha}/(m_{n}+m_{\alpha})$ is the reduced mass, and
$\hat{j}_{\ell}$, $\hat{n}_{\ell}$ are the Riccati-Bessel functions of
the first and second kind, respectively.
The physical scattering phase shift is obtained from the asymptotic value,
\begin{equation}
  \delta_{\ell}(E) = \lim_{r\to\infty}\delta(r;E).
\end{equation}
For the numerical calculations reported here we integrate
Eq.~\eqref{eq:phase_eq} on the grid $r \in [r_{0},\,10.0]$~fm with step
size $\Delta r = 0.01$~fm.
The integration is started at $r_{0} = 0.01$~fm rather than at the origin to avoid the $1/r$ divergence of the Riccati--Bessel function of the second kind, $\hat{n}_{\ell}(kr) \sim (kr)^{-\ell}$ as $r \to 0$; beginning at $r_{0}$ introduces a negligible error since the potential is finite and smoothly varying in this region.
The step size was chosen after a convergence check: halving $\Delta r$
to 0.005~fm changed the computed phase shifts by less than $0.01^{\circ}$ across all energies in the dataset, confirming that the solution is
well-resolved at $\Delta r = 0.01$~fm.
For the present work, we have used the values: $m_{n}=939.565$~MeV$/c^{2}$ and $m_{\alpha}=3727.379$~MeV$/c^{2}$, giving $\mu\approx 750.408$~MeV$/c^{2}$. Laboratory energies from the experimental dataset~\cite{Satchler1968} are
converted to the CM frame via the non-relativistic relation
$E_{\rm cm} = [m_{\alpha}/(m_{n}+m_{\alpha})]\,E_{\rm lab}$.

Two properties of Eq.~\eqref{eq:phase_eq} make it particularly attractive in the present context. First, because it is an initial-value problem, it can be solved by a standard Runge-Kutta integrator with full control over accuracy and without any shooting or boundary-matching procedure. Second, every step of the Runge-Kutta integration is a differentiable operation on the potential values $V(r_{i})$. This means that the gradient of the final phase shift with respect to the neural network parameters $\theta$ can be computed by automatic differentiation~\cite{ChenNODE2018} and passed directly to an optimizer, forming a closed, end-to-end differentiable pipeline from network architecture to observable.

\subsection{Neural Network Architecture}
\label{ssec:arch}
The unknown radial potential is represented as
\begin{equation}
  V_{\theta}(r)
  = \mathcal{N}_{\theta}(r)\;\exp\!\left[-\!\left(\frac{r}{R}\right)^{2}\right],
  \label{eq:Vtheta}
\end{equation}
where $\mathcal{N}_{\theta}(r)$ is a feed-forward network with 
parameters $\theta$, and the Gaussian envelope with scale $R = 3.0$~fm 
enforces the finite-range condition directly at the architectural 
level. At $r = 3R = 9$~fm, well beyond any physically significant 
nuclear force, the envelope already suppresses the potential by a 
factor of $e^{-9} \approx 10^{-4}$, ensuring that $V_{\theta}(r)$ 
vanishes exponentially in the tail region.
We verified that varying $R$ between 2.5 and 4.0~fm does not change the
reconstructed potential or the fitted phase shifts in any meaningful way, confirming that the results are robust to this architectural choice.
The network $\mathcal{N}_{\theta}$ consists of an input layer (a single 
node receiving the rescaled coordinate $r/6$), two hidden layers each 
with 64 nodes and hyperbolic-tangent activation functions, and a single 
linear output node. The rescaling $r/6$ maps the physical range 
$[0, 10]$~fm to approximately $[0, 1.67]$, keeping the network inputs 
within a well-behaved regime of the $\tanh$ activation function and 
avoiding premature saturation during early training.
The network width of 64 nodes per layer was chosen after a small
hyperparameter sweep; narrower networks (32 nodes) showed marginally slower convergence, while wider ones (128 nodes) gave indistinguishable results at greater computational cost.
It is worth being explicit about what happens without the Gaussian factor in Eq.~\eqref{eq:Vtheta}, because the contrast is instructive.
When the raw network output $\mathcal{N}_{\theta}(r)$ is used directly as the potential, three interconnected problems arise.
\begin{enumerate}
\item \textbf{Asymptotic non-vanishing.}
The output of a finite-weight feed-forward network with tanh activations does not in general converge to zero as $r\to\infty$.
It may approach a non-zero constant, oscillate about a baseline, or behave erratically depending on the weight values reached during training. None of these behaviors is consistent with the requirement that nuclear potentials must satisfy $V(r)\to 0$ for $r$ well beyond the range of the strong interaction, a condition that underpins the entire phase-shift formalism. Violating it introduces a spurious long-range Hamiltonian that distorts the scattering boundary conditions.
\item \textbf{Systematic phase-shift errors.}
Because Eq.~\eqref{eq:phase_eq} integrates the potential from $r=r_{0}$ to $r=10$~fm, any non-zero tail of the potential contributes cumulatively to the phase shift throughout the integration range.
In practice, training the unconstrained network leads to interaction strength being spread across large values of $r$ where it should be negligible. The resulting model phase shifts develop a systematic offset from the experimental data that persists even after thousands of training epochs.
The resonant energy dependence, the rapid rise through $90^{\circ}$ that
is the most diagnostic feature of the $P_{3/2}$ channel, is reproduced poorly, and the high-energy behavior is qualitatively wrong.
\item \textbf{Rough loss landscape and slow convergence.}
Without the envelope, the optimizer is free to distribute interaction
strength anywhere along the radial axis.
This dramatically expands the effective dimensionality of the search space and introduces many local minima corresponding to potentials with long-range tails that happen to partially cancel each other's phase-shift contributions.
In repeated trials the training loss fluctuates considerably and the final solutions are neither unique nor physically interpretable.
\end{enumerate}
Multiplying by the Gaussian envelope resolves all three issues simultaneously. Regardless of the values taken by $\theta$ during or after training, $V_{\theta}(r)$ is guaranteed to vanish exponentially for $r\gg R$. This hard constraint is imposed architecturally, before any gradient computation, and it costs nothing in expressive power for the radial range where the interaction is genuinely present.
The optimizer is therefore confined to a subspace of functions that are
physically admissible from the outset, and convergence is fast and robust.
The broader point deserves emphasis: in inverse problems where the physics specifies a boundary or asymptotic condition on the unknown function, encoding that condition into the architecture is strictly superior to encoding it as a soft penalty in the loss.
A soft penalty only discourages violation of the condition; an architectural constraint prevents it absolutely.
For nuclear potentials, where the finite-range condition is as fundamental as conservation of energy, there is no reason to leave it to chance.

\subsection{Loss Function}
\label{ssec:loss}

The model is trained by minimizing the composite loss
\begin{equation}
  \mathcal{L}
  = \mathcal{L}_{\rm data}
  + \lambda_{s}\,\mathcal{L}_{\rm smooth}
  + \lambda_{r}\,\mathcal{L}_{\rm range}
  + \lambda_{\ell}\,\mathcal{L}_{\rm low},
  \label{eq:loss}
\end{equation}
with weighting coefficients $\lambda_{s}=\lambda_{r}=10^{-2}$ and
$\lambda_{\ell}=10^{-3}$.
The individual terms are described below.

\paragraph{Data fidelity}
\begin{equation}
\begin{aligned}
\mathcal{L}_{\rm data}
&= \frac{1}{N}\sum_{i=1}^{N}
w_{i}\bigl[\delta_{\rm model}(E_{i})-\delta_{\rm exp}(E_{i})\bigr]^{2}, \\
w_{i}
&= \frac{1}{1+\lvert E_{i}^{\rm cm}-E_{r}\rvert},
\end{aligned}
\label{eq:Ldata}
\end{equation}
where $E_{r} \approx 0.80$~MeV is the CM resonance energy of the
$P_{3/2}$ channel in $^{5}$He~\cite{Tilley2002}.
The weight $w_{i}$ is largest near the resonance, placing stronger emphasis on the energy region where the phase shift varies most rapidly.
This choice accelerates learning of the most dynamically important feature of the spectrum without discarding any data points.

\paragraph{Smoothness regularization}
\begin{equation}
  \mathcal{L}_{\rm smooth}
  = \frac{1}{N_{r}}\sum_{i}\bigl[\Delta^{2}V_{\theta}(r_{i})\bigr]^{2},
  \label{eq:Lsmooth}
\end{equation}
where $\Delta^{2}$ is the second finite difference evaluated on the
radial grid and the sum runs over all $N_{r}$ grid points.
Phase-shift data alone constrain the integrated effect of the potential
but are largely insensitive to high-frequency oscillations that integrate to zero; this term explicitly suppresses such unphysical oscillations and guides the optimizer toward smooth, physically interpretable solutions.
\paragraph{Finite-range reinforcement}
\begin{equation}
  \mathcal{L}_{\rm range}
  = \frac{1}{N_{>}}\sum_{r_{i}>4\,\rm fm} V_{\theta}(r_{i})^{2},
  \label{eq:Lrange}
\end{equation}
where $N_{>}$ is the number of grid points beyond 4~fm.
Although the Gaussian envelope already guarantees that $V_{\theta}(r)$
decays exponentially, this term provides an additional gradient signal
that actively steers the potential toward zero in the tail region,
accelerating convergence near the cutoff radius.
Its contribution to the total loss is small ($\lambda_{r} = 10^{-2}$),
and it does not replace the architectural enforcement but complements it
during the early stages of training when the network weights are far from their converged values.

\paragraph{Low-energy threshold constraint}
For a $\ell=1$ potential, effective-range theory requires
$\delta_{1}(E)/k^{3}\to -1/a_{1}$ as $k\to 0$,
where $a_{1}$ is the P-wave scattering volume~\cite{Blatt1952}.
We encourage this behavior through
\begin{equation}
  \mathcal{L}_{\rm low}
  = \frac{1}{5}\sum_{i=1}^{5}
    \left[
      \frac{\delta_{\rm model}(E_{i})}{k_{i}^{3}}
      - \overline{\!\left(\frac{\delta}{k^{3}}\right)\!}
    \right]^{2},
  \label{eq:Llow}
\end{equation}
summed over the five lowest energies in the dataset, where
$\overline{(\delta/k^{3})}$ is the mean of $\delta_{\rm model}/k^{3}$
over those same five points.
Rather than enforcing a specific value of the scattering volume, this
term penalizes departures from a flat $\delta/k^{3}$ ratio at threshold,
which is the qualitative content of effective-range theory for P-waves.
In this sense, $\mathcal{L}_{\rm low}$ acts as a smoothness constraint
on the low-energy tail of the phase-shift curve, stabilizing training
without requiring prior knowledge of $a_{1}$.

\subsection{Training Procedure}
\begin{figure}[htbp]
  \centering
  \includegraphics[width=0.92\textwidth]{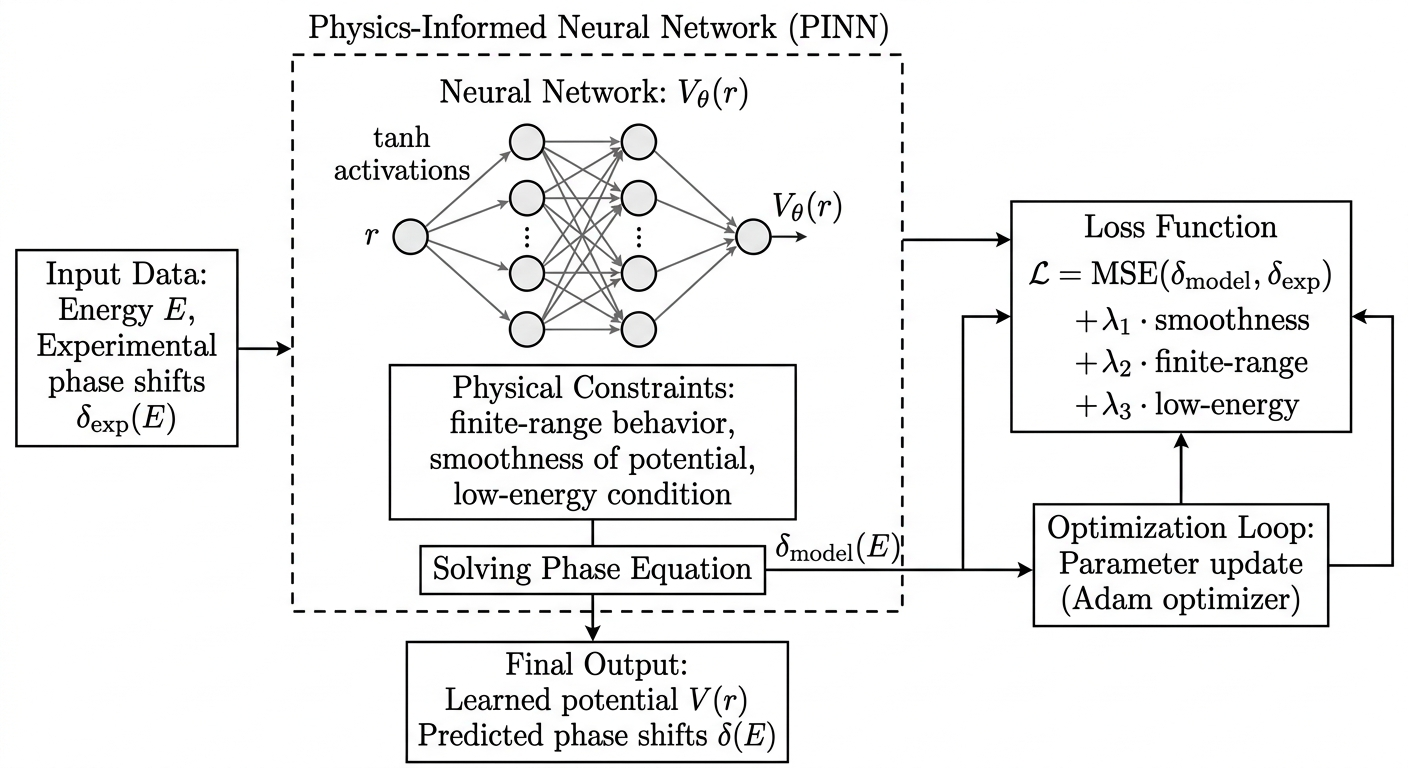}
  \caption{Schematic of the PINNs framework.}
  \label{fig:flowchart}
\end{figure}

The full dataset of 22 phase-shift values from Ref.~\cite{Satchler1968} is used in every update step (full-batch gradient descent).
Network parameters are updated using the Adam optimizer~\cite{Kingma2015} at a fixed learning rate of $10^{-4}$.
Training runs for 6000 epochs; we verified that all loss components plateau well before this point and that extending training to 10000 epochs produces no measurable change in the reconstructed potential or phase shifts. The random seed is fixed to 2 for the primary reported result. The sensitivity of the solution to this choice is quantified explicitly in the robustness tests described below.
All computations are carried out in PyTorch using automatic differentiation to propagate gradients through the Runge-Kutta integration~\cite{ChenNODE2018}. 
The schematic of the full training pipeline is shown in Fig.~\ref{fig:flowchart}.
The source code used in this work, including the PINN architecture, the Runge-Kutta variable-phase solver, and the training pipeline, is available at GitHub \cite{awasthi2026pinn}.
\subsection{Robustness Tests}
To assess the reliability of the learned inverse potential, we carried out three types of test.
\begin{enumerate}
  \item \textit{Initialization sensitivity.}
        The model is retrained from ten different random initializations. If the reconstructed potentials are nearly identical across all runs, the optimization landscape is effectively unimodal and the solution can be trusted to be globally, rather than locally, optimal.
  \item \textit{Epoch sensitivity.}
Training is stopped at 500, 1000, 2000, 4000, 6000 and 10000 epochs to
map out how quickly and smoothly the solution converges toward its
final form.
   \item \textit{Leave-one-out (LOO) analysis.}
The model is retrained 22 times, each time removing one data point from the training set~\cite{Satchler1968}. The spread of the resulting ensemble of potentials quantifies the sensitivity of the inverse solution to individual measurements.
\end{enumerate}

\section{Results and Discussion}
\label{sec:results}

\subsection{Convergence of Training}

\begin{figure}[htbp]
  \centering
  \includegraphics[width=0.6\textwidth]{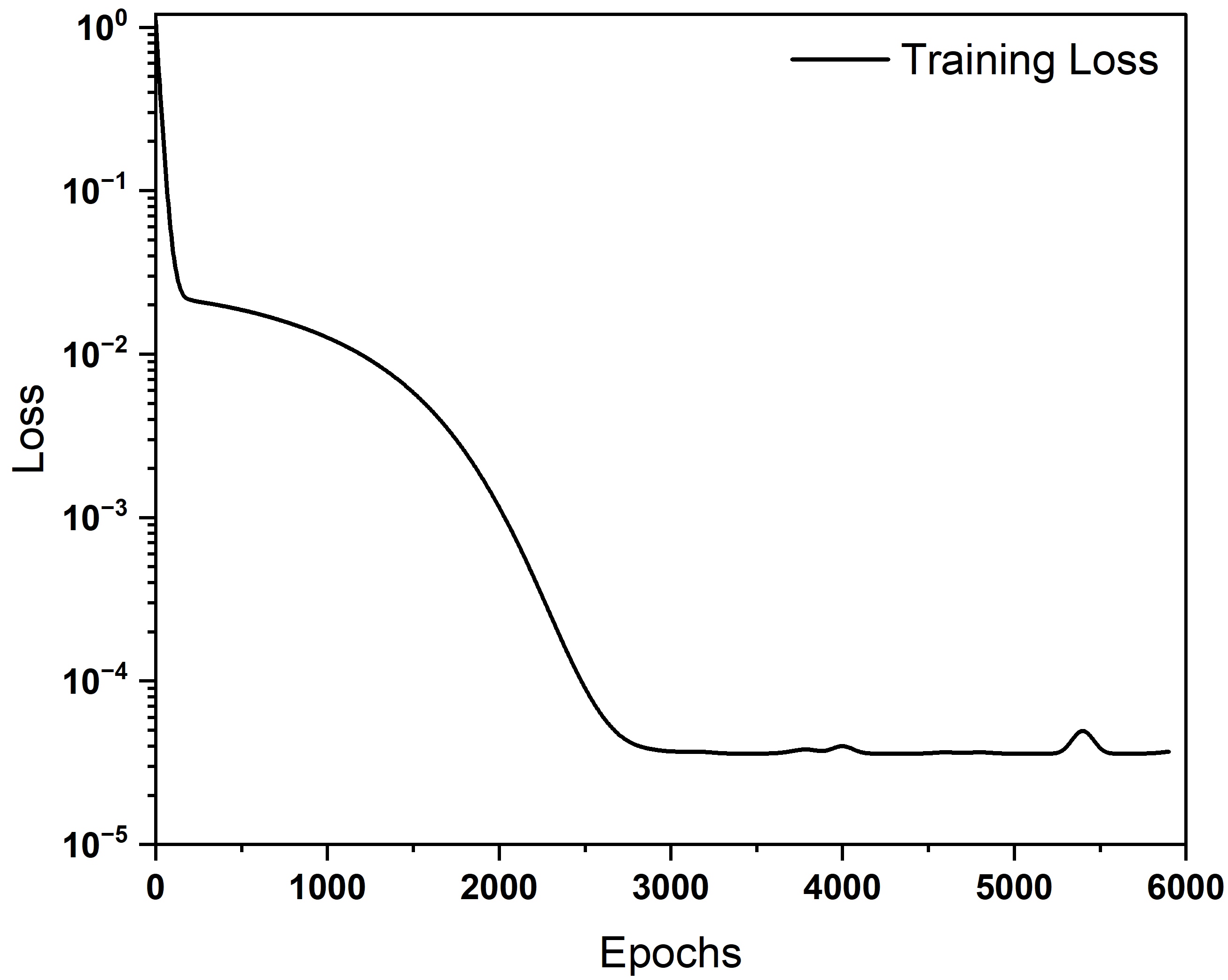}
  \caption{Training loss on a logarithmic scale as a function of
           epoch.  The loss decreases by four orders of magnitude and
           stabilises near $3\times10^{-4}$, indicating well-conditioned
           optimisation with no signs of overfitting.}
  \label{fig:loss}
\end{figure}

The evolution of the training loss as a function of epoch is shown in
Fig.~\ref{fig:loss}.  Starting from values near unity, $\mathcal{L}$
drops by four orders of magnitude within the first 3\,000 epochs and
subsequently settles near $3\times10^{-4}$.  The descent is smooth
and monotonically decreasing for most of the training run; a minor
transient bump visible near epoch 5\,300 arises from a brief
competition between the data-fidelity and smoothness gradients and
resolves automatically within a few hundred steps, leaving no lasting
imprint on the converged solution.

That the loss should converge so cleanly is, to some extent, a
direct consequence of the architectural choice made for the potential
ansatz.  Ill-posed inverse problems are known to harbour intricate
loss landscapes populated by spurious valleys and saddle
points~\cite{adler2017solving}.  By enforcing a Gaussian envelope on the
network output, unphysical long-range degrees of freedom are
effectively eliminated; these are precisely the directions in
parameter space responsible for the topological complexity just
mentioned.  The clean convergence observed here therefore reflects,
in a transparent way, the regularising benefit of this built-in
physical constraint.

\subsection{Reconstructed Potential}

\subsubsection*{Central potential}

The learned central potential $V_{\theta}(r)$ is shown in the
left side of Fig.~\ref{fig:potential}.  The profile is purely
attractive, reaching a minimum of approximately $-60.47$~MeV and
returning smoothly to zero beyond $r \approx 4$~fm; no repulsive
core is present at short distances.

This shape is physically reasonable for the $P_{3/2}$ partial
wave.  In channels with $\ell \geq 1$, Pauli blocking between the
projectile neutron and the constituent nucleons of the $\alpha$
cluster is considerably weaker than in $S$-wave
scattering, so the dominant contribution to the
effective neutron-$\alpha$ interaction is attractive throughout the
nuclear interior.  Additionally, the $\alpha$ particle is a tightly
bound, spin-zero composite object, the neutron couples to it as a
whole rather than to the individual nucleons.  Averaging over the
intra-cluster degrees of freedom naturally washes out the hard core
present in the bare nucleon-nucleon interaction, yielding the soft,
purely attractive profile seen here.

The absence of oscillations in $V_{\theta}(r)$ confirms that the
smoothness penalty $\mathcal{L}_{\rm smooth}$ is functioning as
intended, while the finite-range Gaussian envelope guarantees the
correct asymptotic behaviour without introducing tension in the
optimisation.

\subsubsection*{Effective potential including the centrifugal term}

\begin{figure}[htbp]
  \centering
  \includegraphics[width=\textwidth]{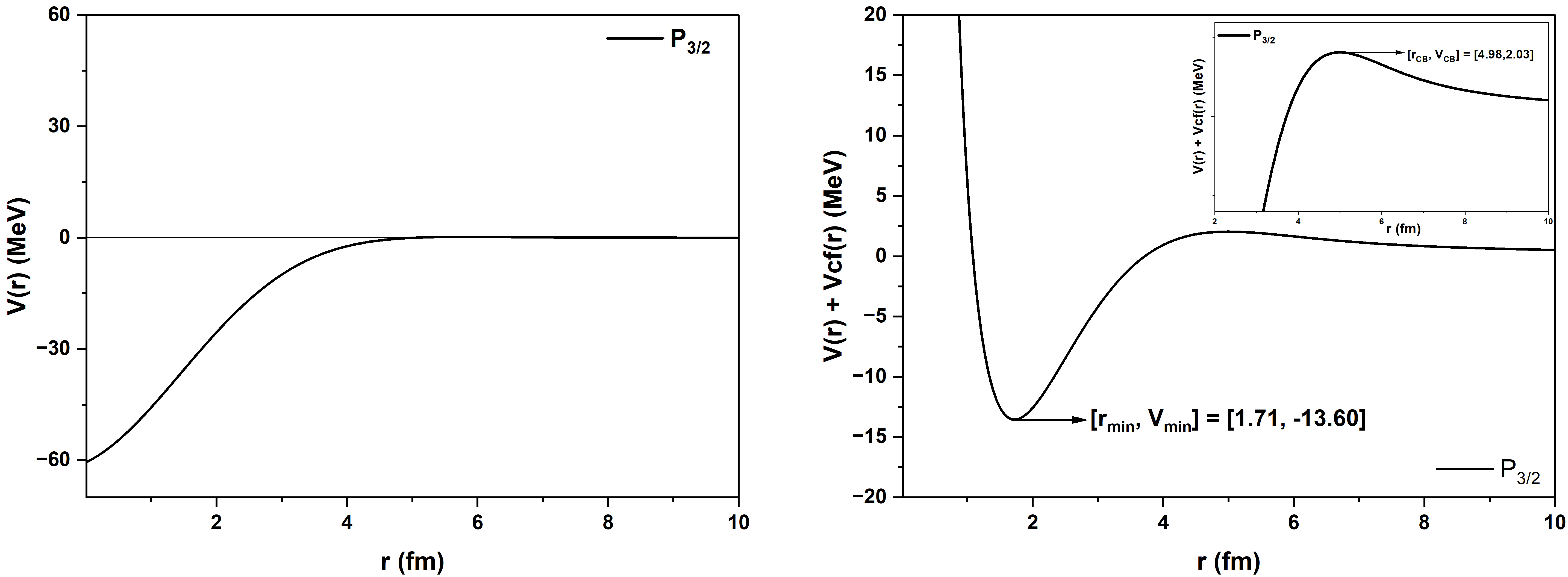}
  \caption{Left: learned central potential $V_{\theta}(r)$ for the
           $P_{3/2}$ channel.  The profile is purely attractive and
           smooth, vanishing beyond $\approx 4$~fm.  Right: effective
           potential $V_{\rm eff}(r) = V_{\theta}(r)+V_{\rm cf}(r)$,
           displaying the barrier--well structure responsible for the
           $P_{3/2}$ resonance.  Inset: enlarged view of the potential
           minimum and centrifugal barrier peak.}
  \label{fig:potential}
\end{figure}

Upon adding the centrifugal term to the learned central
potential, the effective potential takes the form
\begin{equation}
  V_{\rm eff}(r)
  = V_{\theta}(r)
    + \frac{\ell(\ell+1)\hbar^{2}}{2\mu r^{2}},
  \label{eq:Veff}
\end{equation}
and is shown in the right side of Fig.~\ref{fig:potential}.  The
purely attractive central potential is transformed into a pronounced
barrier-well structure: a potential well of depth approximately
$13.6$~MeV centered near $r = 1.71$~fm, and a centrifugal
barrier reaching a maximum of $= 2.03$~MeV near
$r = 4.98$~fm.

This topology provides a direct, intuitive explanation for the
observed resonance.  A neutron incident on the $\alpha$ particle
with energy near the barrier height can become temporarily trapped
inside the attractive well, forming a quasi-bound state that persists
for a finite lifetime before tunnelling through the barrier and
escaping to the continuum.  During the trapping interval the phase
shift passes rapidly through $90^{\circ}$, which is the kinematic
signature of a resonance.  The extracted potential parameters are given as 
\begin{equation*}
  [r_{\rm min},\;V_{\rm min}] = [1.71~\text{fm},\;{-13.60}~\text{MeV}],
\end{equation*}
\begin{equation*}
  [r_{\rm CB},\;V_{\rm CB}] = [4.98~\text{fm},\;{+2.03}~\text{MeV}],
\end{equation*}
%

\begin{figure}[htbp]
  \centering
  \includegraphics[width=0.6\textwidth]{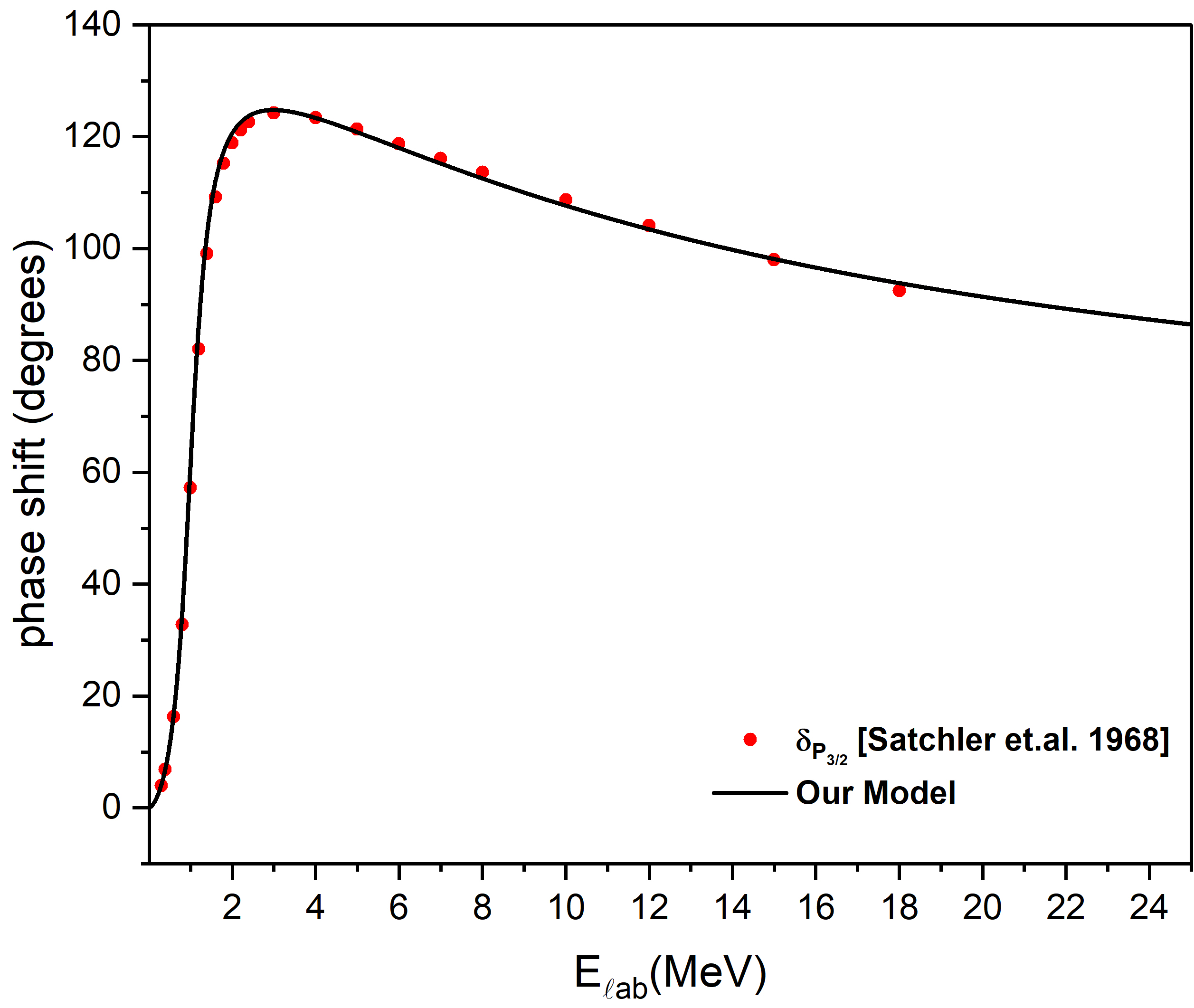}
  \caption{$P_{3/2}$ phase shifts for $n$-$\alpha$ scattering as a
           function of laboratory energy $E_{n}$.  The solid line is the
           present PINN result; filled circles are the expected values
           of Satchler~\textit{et~al.}~\cite{Satchler1968}.  The
           shaded bands indicate the extrapolation regions outside the
           training interval $0.3$--$18$~MeV.}
  \label{fig:sps}
\end{figure}

Using the reconstructed potential, the $P_{3/2}$ phase shifts have
been computed over the energy range $0.01$-$25$~MeV and are compared
with the data of Satchler~\textit{et~al.}~\cite{Satchler1968} in
Fig.~\ref{fig:sps}.  The network was trained exclusively on the
interval $0.3$-$18$~MeV, so the phase shifts below $0.3$~MeV and
above $18$~MeV constitute  model predictions extending into
both the low- and high-energy regimes.  Agreement with the reference
data is good across the entire evaluated range.  In detail, the model
reproduces the rapid rise through $90^{\circ}$ near $0.9$~MeV, the
broad plateau between $120^{\circ}$ and $125^{\circ}$ from roughly
$2$ to $4$~MeV, and the gradual decline to about $93^{\circ}$ at
$18$~MeV.  The smooth, continuous character of the predicted curve
confirms that the network generalises between the training energies
without overfitting individual data points.

\subsection{Residual Analysis}

\begin{figure}[htbp]
  \centering
  \includegraphics[width=0.6\textwidth]{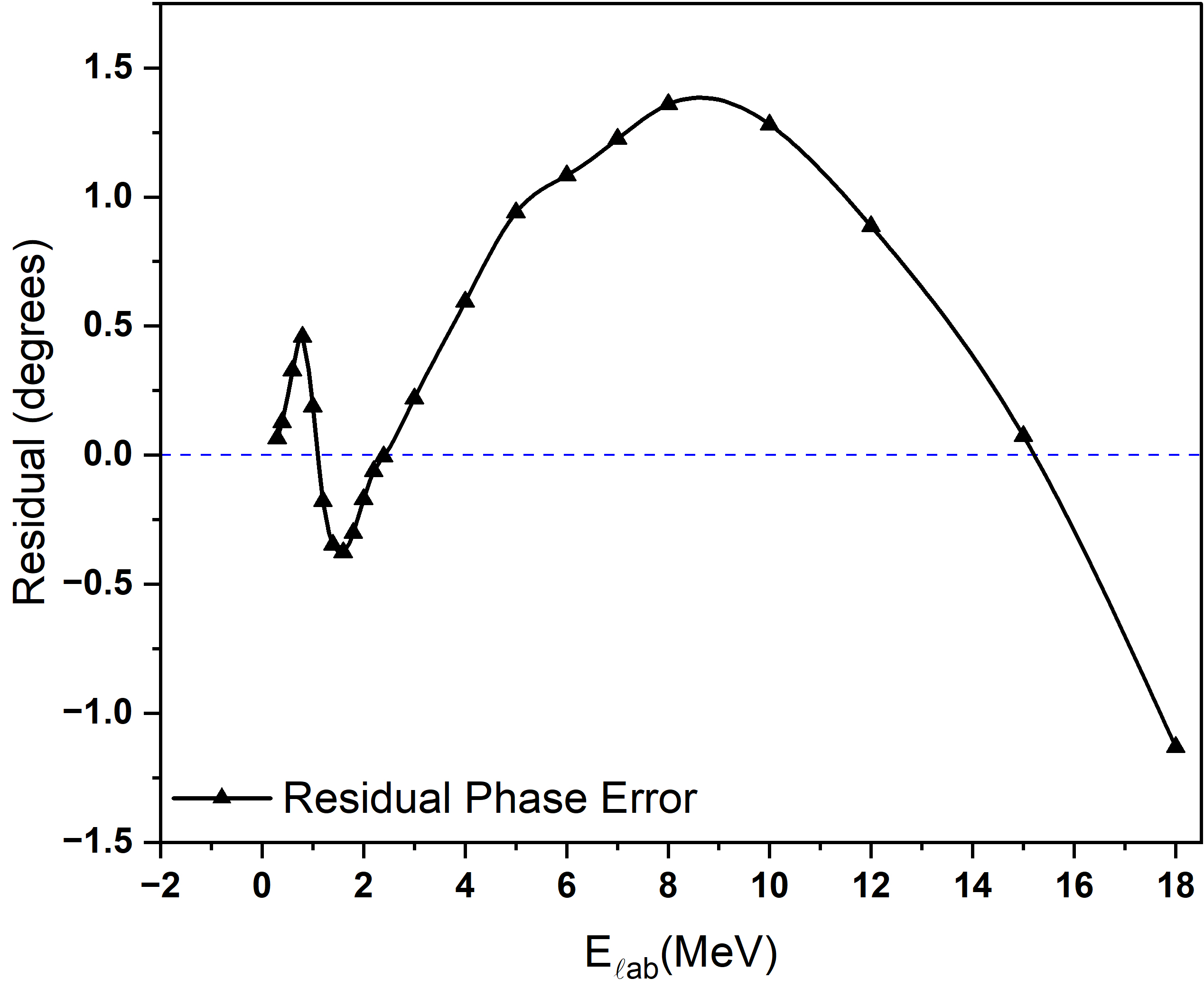}
  \caption{Residual $\Delta\delta = \delta_{\rm model} - \delta_{\rm exp}$
           as a function of laboratory energy.  All 22 residuals lie
           within $\pm 1.4^{\circ}$; the mean-square value is
           $0.5^{\circ}$.  The absence of any systematic trend
           confirms unbiased reproduction of the phase shifts.}
  \label{fig:residual}
\end{figure}

The pointwise residual $\Delta\delta = \delta_{\rm model} -
\delta_{\rm exp}$, plotted against laboratory energy in
Fig.~\ref{fig:residual}, lies within $\pm 1.4^{\circ}$ at all 22
training points.  The mean-square residual over the full set
is $\Delta\delta_{\rm mse} = 0.5^{\circ}$. The distribution is roughly symmetric about zero and shows no
discernible trend with energy, which is the expected signature of a
model fitting the data uniformly rather than struggling in any
particular energy region.
The somewhat larger residuals in the neighbourhood of
$E_{\rm lab} \approx 8$-$10$~MeV are physically understandable,
beyond the resonance peak the phase shift varies slowly with energy,
reducing the effective gradient signal in the loss function and
leading to correspondingly less tight local convergence.  No
systematic bias is implied.

\subsection{Partial Cross Section and Resonance Parameters}

\begin{figure}[htbp]
  \centering
  \includegraphics[width=0.60\textwidth]{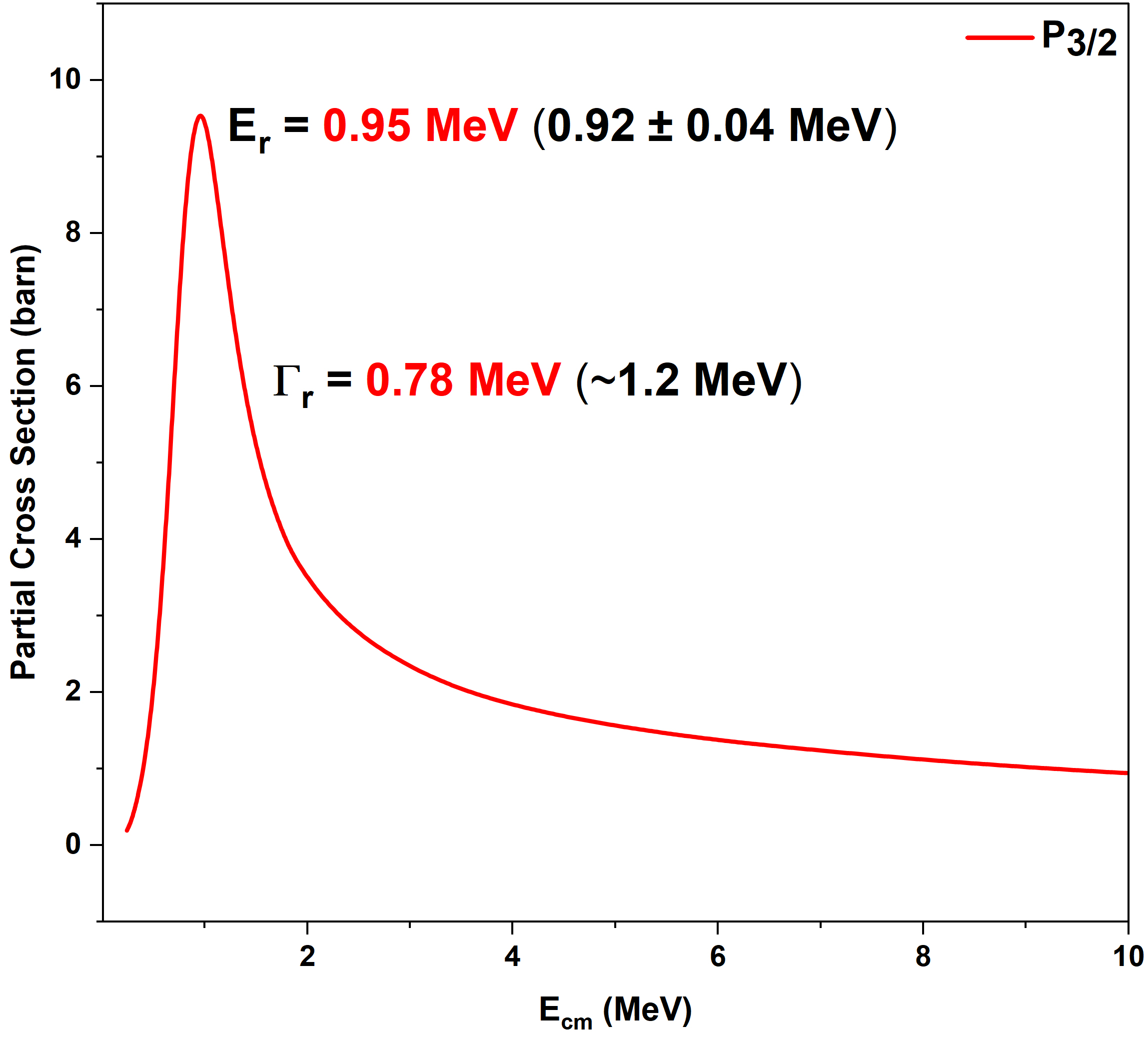}
  \caption{Partial-wave cross section $\sigma_{1}(E)$ for the
           $P_{3/2}$ channel of
           $^{4}{\rm He}(n,n)^{4}{\rm He}$ as a function of
           centre-of-mass energy.  The resonance peak at
           $E_{\rm cm} = 0.95$~MeV yields $\Gamma = 0.78$~MeV;
           experimental values are shown in parentheses.}
  \label{fig:scs}
\end{figure}

From the computed phase shifts, the partial-wave cross
section for $\ell = 1$ is obtained from
\begin{equation}
  \sigma_{1}(E)
  = \frac{4\pi}{k^{2}}\,(2\ell+1)\sin^{2}\!\delta_{1}(E),
  \label{eq:xsec}
\end{equation}
and is shown in Fig.~\ref{fig:scs} as a function of the
centre-of-mass energy $E_{\rm cm}$ for the
$^{4}{\rm He}(n,n)^{4}{\rm He}$ reaction.

A well-defined resonance peak appears at $E_{\rm cm} = 0.95$~MeV
with a width $\Gamma = 0.78$~MeV.  
These values are identified with

A quantitative comparison with R-matrix results and experiment is
presented in Table~\ref{tab:resonance_comparison}.  The resonance
energy $E_{\rm cm} = 0.95$~MeV agrees with the experimental value of
$0.92 \pm 0.04$~MeV \cite{Ajzenberg1979} to within $0.03$~MeV ($\approx 3\%$). The
calculated resonance width, $\Gamma = 0.78$~MeV, falls between the
R-matrix value of $0.64$~MeV (evaluated at channel radius
$a = 3.3$~fm)~\cite{Borsaru1977} and the experimentally measured
FWHM of $\approx 1.2$~MeV from the total cross
section~\cite{Ajzenberg1979}.  As is well known, the
R-matrix pole width depends sensitively on the channel radius, with
smaller values of $a$ yielding narrower resonances~\cite{Barker1985};
the present result sits naturally between the two reference values.

\begin{table*}[htbp]
\centering
\caption{Comparison of $P_{3/2}$ resonance parameters for
         $^{4}{\rm He}(n,n)^{4}{\rm He}$ obtained from the present
         model, R-matrix analysis~\cite{Borsaru1977}, and
         experiment~\cite{Ajzenberg1979}.}
\label{tab:resonance_comparison}
\begin{tabular}{lccc}
\hline\hline
\textbf{Parameter}
  & \textbf{Present model}
  & \textbf{R-matrix}~\cite{Borsaru1977}
  & \textbf{Experiment}~\cite{Ajzenberg1979} \\
\hline
Resonance energy $E_{\rm cm}$ (MeV) & $0.95$       & $0.77$  & $0.92\pm0.04$ \\
Resonance width $\Gamma$ (MeV)      & $0.78$       & $0.64$  & ${\sim}1.2$~(FWHM) \\
\hline\hline
\end{tabular}
\end{table*}
\subsection{Effective-Range Parameters}

At low energies the $P$-wave phase shift obeys the effective-range
expansion~\cite{Blatt1952,Arndt1973},
\begin{equation}
  k^{3}\cot\delta_{1}
  = -\frac{1}{a}
    + \frac{1}{2}\,r\,k^{2}
    - \frac{1}{4}\,P\,k^{4} + \cdots,
  \label{eq:ere}
\end{equation}
where $a$ is the scattering volume, $r$ the effective range, and $P$
the shape parameter.

Equation~\eqref{eq:ere} was fitted to the five lowest-energy model
phase shifts in the training set, spanning the range
$E_{\rm cm} \approx 0.3$-$1.5$~MeV, where truncation of the
expansion at order $k^{4}$ is well justified.  In practice,
Eq.~\eqref{eq:ere} is a polynomial in $k^{2}$,
\begin{equation}
  k^{3}\cot\delta_{1} = A + B\,k^{2} + C\,k^{4} + \cdots,
  \label{eq:ere_poly}
\end{equation}
with $a = -1/A$, $r = 2B$, and $P = -4C$.  The coefficients
$A$, $B$, $C$ were determined by a standard nonlinear least-squares
fit, and their statistical uncertainties $\sigma_{A}$, $\sigma_{B}$,
$\sigma_{C}$ were obtained from the diagonal elements of the
fit covariance matrix.  The uncertainties on the physical parameters
were then propagated analytically:
\begin{equation}
  \sigma_{a} = \frac{\sigma_{A}}{A^{2}}, \quad
  \sigma_{r} = 2\,\sigma_{B}, \quad
  \sigma_{P} = 4\,\sigma_{C}.
  \label{eq:ere_errors}
\end{equation}
The resulting parameters are presented in Table~\ref{tab:ere}
alongside the analysis of Arndt~\textit{et~al.}~\cite{Arndt1973}.

The scattering volume and effective range reproduce the reference
values to within $1.2\%$ and $1.7\%$, respectively, a level of
agreement that is quite satisfying given that neither quantity was
directly constrained during training.  The shape parameter $P$
shows a larger deviation of $\approx 12\%$, as $P$ is sensitive to higher powers of $k$ and therefore probes short-distance features of the potential that are less tightly constrained by the few lowest-energy data points entering the fit.

\begin{table*}[htbp]
  \centering
  \caption{$P$-wave effective-range parameters from the PINN model
           compared with the analysis of Arndt~\textit{et~al.}\
           \cite{Arndt1973}.  Uncertainties are one-standard-deviation
           estimates obtained by propagating the covariance matrix of
           the least-squares polynomial fit to
           $k^{3}\cot\delta_{1}$ via Eq.~\eqref{eq:ere_errors}.}
  \label{tab:ere}
  \begin{tabular}{ccc}
    \toprule
    Parameter
      & PINNs
      & Arndt~\textit{et~al.}~\cite{Arndt1973} \\
    \midrule
    Scattering volume $a$\,(fm$^{3}$)
      & $-62.182 \pm 0.940$
      & $-62.951 \pm 0.003$ \\
    Effective range $r$\,(fm)
      & $-0.867  \pm 0.007$
      & $-0.8819 \pm 0.0011$ \\
    Shape parameter $P$\,(fm$^{5}$)
      & $-2.653  \pm 0.032$
      & $-3.002  \pm 0.062$ \\
    \bottomrule
  \end{tabular}
\end{table*}
\subsection{Robustness Analysis}
Three separate tests were carried out to verify that the reconstructed
potential is not an artefact of particular algorithmic choices.
\paragraph{Initialization sensitivity:}
The network was retrained from ten independent random initializations
(\texttt{randomseed(3)}, \texttt{randomseed(10)}, \texttt{randomseed(20)}, \texttt{randomseed(30)}, \texttt{randomseed(40)}, \texttt{randomseed(100)}, \texttt{randomseed(200)}, \texttt{randomseed(1000)}, \texttt{randomseed(2000)}
and \texttt{randomseed(2026)}).
In every case the reconstructed potential $V_{eff}(r)$ was indistinguishable
from the baseline result to within plotting precision.
This consistency across seeds suggests that the loss landscape encountered during training is effectively unimodal in the relevant parameter region, and that the solution reported here is not a lucky outcome of a single favourable starting point.

\paragraph{Epoch sensitivity:}
Training was halted at 500, 1000, 2000, 4000, 6000 and 10000 epochs to examine convergence behaviour. Stopping before 3\,000 epochs produced noticeably elevated residuals and a potential that had not yet settled to its final shape. Between 4\,000 and 6\,000 epochs the curve changed only marginally, and extending training beyond 6\,000 epochs brought no measurable improvement in either the fit quality or the recovered potential. On this basis, 6\,000 epochs at a learning rate of $10^{-4}$ was adopted as the training budget throughout this work.

\begin{figure}[htbp]
  \centering
  \includegraphics[width=0.6\textwidth]{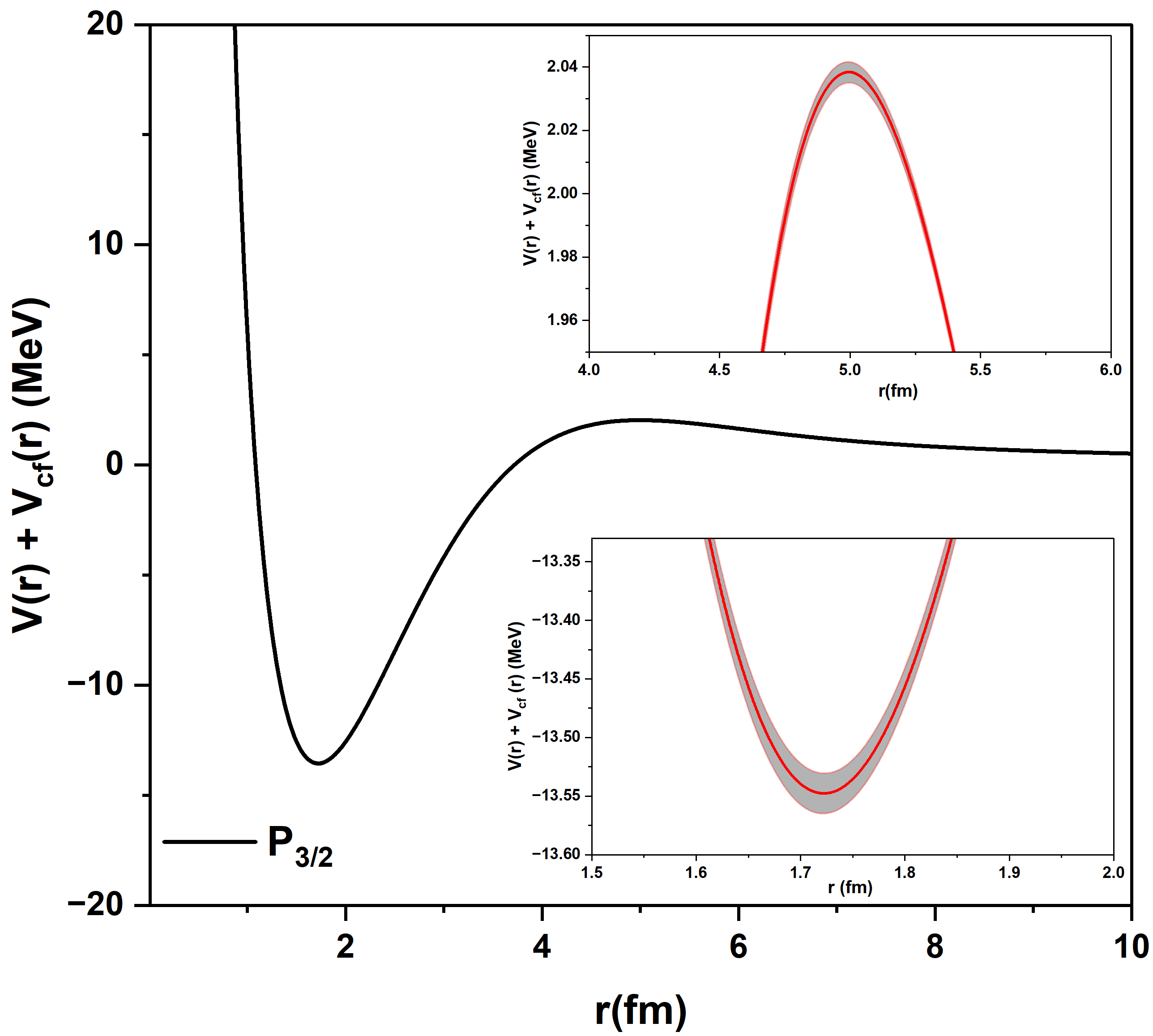}
  \caption{Leave-one-out robustness analysis of the reconstructed
effective potential $V_{\rm eff}(r)$ for the $P_{3/2}$ partial
wave. Insets show enlarged views near the potential minimum
($r \approx 1.7$~fm) and barrier peak ($r \approx 5.0$~fm); the
red line is the LOO mean and the grey band the $\pm1\sigma$ spread.
The sub-hundredths-of-an-MeV bandwidth confirms that the
reconstruction is not sensitive to any individual data point.}
\label{fig:loo}
\end{figure}

\paragraph{Leave-one-out analysis.}
The most stringent of the three tests is the LOO ensemble, displayed in
Fig.~\ref{fig:loo}.
The model was retrained 22 times, each time with one data point withheld, and the resulting potentials were overlaid to form a spread band.
The $\pm1\sigma$ band is remarkably narrow across the full range of $r$,
amounting to no more than a few hundredths of an MeV even at the two
physically sensitive features: the potential minimum near $r = 1.7$~fm and
the barrier peak near $r = 5.0$~fm (see insets of Fig.~\ref{fig:loo}).
The implication is clear,-no single data point is controlling the shape of the reconstruction.
This is a non-trivial finding given the ill-posed character of the inverse scattering problem~\cite{Tarantola2005}, and it provides tangible support for the reliability of the PINN framework in this context.
The fact that all randomly initialized runs fall inside the LOO band
further reinforces the earlier observation that the optimization converges robustly to a unique solution.

\section{Conclusion}
\label{sec:conclusion}
We have developed a physics-informed neural network framework for the
inverse scattering problem in nuclear physics and applied it to the
$P_{3/2}$ partial wave of neutron-$\alpha$ elastic scattering.
The central design choice, multiplying the network output by a
Gaussian envelope to enforce finite range at the architectural
level, proved essential for obtaining physically meaningful results.
Without this constraint, training consistently failed to reproduce the
experimental phase shifts regardless of the number of epochs, a finding
that we regard as a practically important lesson for future PINN
applications to nuclear inverse problems: physical boundary conditions
should be built into the network architecture rather than delegated to
soft penalty terms.
\\With the envelope in place, the training loss converged smoothly to
$3\times10^{-4}$ within 6\,000 epochs and the reconstructed central
potential is purely attractive, smooth, and correctly finite-ranged,
with a well depth of approximately $-60.47$~MeV. Adding the centrifugal
term produces a barrier-well structure with a well depth of
$-13.60$~MeV at $r = 1.71$~fm and a barrier height of $+2.03$~MeV at
$r = 4.98$~fm, which provides a transparent physical picture of the
$^{5}$He resonance as a quasi-bound state formed by temporary
trapping behind the centrifugal barrier. The extracted resonance
parameters, $E_{r} = 0.95$~MeV and $\Gamma = 0.78$~MeV, are in good
agreement with the experimental values \cite{Ajzenberg1979}. The P-wave effective-range parameters reproduce
the reference values of Arndt~\textit{et~al.}~\cite{Arndt1973} to
within $1.2\%$ and $1.7\%$ for the scattering volume and effective
range, respectively, despite neither quantity being directly
constrained during training.

Robustness tests lend further credibility to these results.
Retraining from multiple random initializations yields potentials that
are indistinguishable at the level of plotting precision, indicating
that the optimization landscape is effectively unimodal. The
leave-one-out analysis, which is the most demanding of the three
tests, shows that the $\pm1\sigma$ ensemble band is only a few
hundredths of an MeV wide even at the most physically sensitive
features of the effective potential. For an ill-posed inverse problem
of this kind, such stability is not automatic and provides genuine
evidence that the reconstructed potential reflects the physics of the
data rather than any particular algorithmic accident.

Taken together, these results establish the PINN framework as a
reliable and interpretable tool for potential reconstruction from
nuclear scattering data, combining the flexibility of neural networks
with the stability afforded by hard physical constraints.
Extensions of this work include the simultaneous treatment of
both $P$-wave channels to recover the spin-orbit interaction, the
application to heavier nucleon-nucleus systems where phenomenological
potentials are less well constrained, and a more systematic treatment
of experimental uncertainties through ensemble training with noise
augmentation. We intend to pursue these directions in subsequent work.
\section{Acknowledgments}
A. Awasthi acknowledges the financial support provided by the Department of Science and Technology (DST), Government of India vide Grant No. DST/INSPIRE Fellowship/2020/IF200538.


\begin{thebibliography}{99}
\bibitem{arrington2022progress} J. Arrington, N. Fomin, and A. Schmidt, Progress in understanding short-range structure in nuclei: an experimental perspective, Annual Review of Nuclear and Particle Science 72, 307 (2022).
\bibitem{Newton1982}R. G. Newton, Scattering Theory of Waves and Particles, 2nd ed. (Springer, New York, 1982).
\bibitem{Chadan1977}K. Chadan and P. C. Sabatier, Inverse Problems in
Quantum Scattering Theory (Springer, New York, 1977).
\bibitem{Machleidt2011}R. Machleidt and D. R. Entem, Chiral effective field theory and nuclear forces, Phys. Rep. 503, 1 (2011).
\bibitem{Satchler1968}G. R. Satchler, L. W. Owen, A. J. Elwyn, G. L. Morgan,and R. L. Walter, An optical model for the scattering of
nucleons from $^{4}He$ at energies below 20 MeV, Nucl. Phys. A 112, 1 (1968)
\bibitem{Tilley2002}D. R. Tilley, C. M. Cheves, J. L. Godwin, G. M. Hale, H. M. Hofmann, J. H. Kelley, C. G. Sheu, and H. R. Weller, Energy levels of light nuclei A = 5, 6, 7, Nucl.Phys. A 708, 3 (2002)
\bibitem{Nollett2007}K. M. Nollett, S. C. Pieper, R. B. Wiringa, J. Carlson, and G. M. Hale, Quantum Monte Carlo calculations of
neutron-$\alpha$ scattering, Phys. Rev. Lett. 99, 022502 (2007).
\bibitem{Lynn2016}J. E. Lynn, I. Tews, J. Carlson, S. Gandolfi, A. Gezerlis, K. E. Schmidt, and A. Schwenk, Chiral three-nucleon interactions in light nuclei, neutron-$\alpha$ scattering, and neutron matter, Phys. Rev. Lett. 116, 062501 (2016).
\bibitem{Hodgson1994}P. E. Hodgson, The Nucleon Optical Model (World Scientific, Singapore, 1994).
\bibitem{Wigner1947}E. P. Wigner and L. Eisenbud, Higher angular momenta
and long range interaction in resonance reactions, Phys.
Rev. 72, 29 (1947).
\bibitem{Descouvemont2010}P. Descouvemont and D. Baye, The R-matrix theory, Rep. Prog. Phys. 73, 036301 (2010).
\bibitem{awasthi2024high} A. Awasthi, A. Sharma, I. Kant, and O. S. K. S. Sastri, High-precision inverse potentials for neutron-proton scat-
tering using piece-wise smooth Morse functions, Chin.
Phys. C 48, 104104 (2024).
\bibitem{awasthi2025genetic}A. Awasthi, A. Sharma, Barbie, I. Kant, and O. S. K. S. Sastri, Genetic-algorithm-based inverse potentials for res-
onant states of $\alpha-^{12}C$ using the variable phase approach,
Phys. Rev. C 112, 054604 (2025).
\bibitem{awasthi2025genetic1}A. Awasthi, A. Sharma, I. Kant, M. R. G. Kumar, and O. S. K. S. Sastri, Genetic algorithm-based in-
verse optimization of interaction potential for nucleon- deuteron scattering below break-up threshold, Comput. Phys. Commun. , 109800 (2025).
\bibitem{sharma2025novel}A. Sharma, A. Awasthi, and O. S. K. S. Sastri, A novel computational approach for study of proton-proton scattering, Sci. Rep. 15, 33764 (2025)
\bibitem{sharma2026genetic}A. Sharma, A. Awasthi, J. Sharma, I. Kant, M. R. G. Kumar, and O. S. K. S. Sastri, Genetic algorithm approach
to study low-energy alpha-deuteron elastic scattering using the phase function method, Prog. Theor. Exp. Phys. 2026, 023D01 (2026).
\bibitem{Boehnlein2022}A. Boehnlein, M. Diefenthaler, N. Sato, M. Schram, V. Ziegler, C. Fanelli, M. Hjorth-Jensen, T. Horn, M. P. Kuchera, D. Lee, W. Nazarewicz, P. Ostroumov, K. Orginos, A. Poon, X.-N. Wang, A. Scheinker, M. S. Smith, and L.-G. Pang, Colloquium: Machine learning in nuclear physics, Rev. Mod. Phys. 94, 031003 (2022).
\bibitem{Neufcourt2019} L. Neufcourt, Y. Cao, W. Nazarewicz, E. Olsen, and F. Viens, Neutron drip line in the Ca region from Bayesian model averaging, Phys. Rev. Lett. 122, 062502 (2019)
\bibitem{Utama2016} R. Utama, J. Piekarewicz, and H. B. Prosper, Nuclear
mass predictions for the crustal composition of neutron stars: A Bayesian neural network approach, Phys. Rev. C 93, 014311 (2016).
\bibitem{Negoita2019}G. A. Negoita, J. P. Vary, G. R. Luecke, P. Maris, A. M. Shirokov, I. J. Shin, Y. Kim, E. G. Lee, J. W. Rehr,
J. Rotureau, and K. A. Wendt, Deep learning: Extrapolation tool for ab initio nuclear theory, Phys. Rev. C 99, 054308 (2019).
\bibitem{Lovell2017}A. E. Lovell, F. M. Nunes, J. Sarich, and S. M. Wild, Quantifying uncertainties in optical model parameters,
Phys. Rev. C 95, 024611 (2017).
\bibitem{Negoita2018}F. Negoita et al., Deep learning techniques for nuclear astrophysics, Nature Communications 9, 1 (2018).
\bibitem{Adams2021}C. Adams, G. Carleo, A. Lovato, and N. Rocco, Variational Monte Carlo calculations of A $\leq$ 4 nuclei with an artificial neural-network correlator ansatz, Phys. Rev. Lett. 127, 022502 (2021).
\bibitem{Raissi2019}M. Raissi, P. Perdikaris, and G. E. Karniadakis, Physics- informed neural networks: A deep learning framework for
solving forward and inverse problems involving nonlinear partial differential equations, J. Comput. Phys. 378, 686 (2019)
\bibitem{Lagaris1998}. E. Lagaris, A. Likas, and D. I. Fotiadis, Artificial neural networks for solving ordinary and partial differential
equations, IEEE Trans. Neural Netw. 9, 987 (1998).
\bibitem{Cuomo2022}S. Cuomo, V. S. di Cola, F. Giampaolo, G. Mascolini,
M. Raissi, and F. Piccialli, Scientific machine learning through physics-informed neural networks: Where we are and what’s next, J. Sci. Comput. 92, 88 (2022).
\bibitem{Calogero1967}Calogero, Variable Phase Approach to Potential Scattering (Academic Press, New York, 1967).
\bibitem{Calogero1963}F. Calogero, Variable phase approach to potential scattering, Il Nuovo Cimento 27, 261 (1963).
\bibitem{palov2021vpa}A. Palov and G. G. Balint-Kurti, Vpa: computer program for the computation of the phase shift in atom–
atom potential scattering using the variable phase approach, Computer Physics Communications 263, 107895 (2021).
\bibitem{ChenNODE2018}R. T. Q. Chen, Y. Rubanova, J. Bettencourt, and D. Duvenaud, Neural ordinary differential equations, in Advances in Neural Information Processing Systems, Vol. 31 (2018) arXiv:1806.07366.
\bibitem{Blatt1952}J. M. Blatt and V. F. Weisskopf, Theoretical Nuclear
Physics (Wiley, New York, 1952).
\bibitem{Kingma2015} D. P. Kingma and J. Ba, Adam: A method for stochastic optimization, in 3rd International Conference on Learn-
ing Representations (ICLR) (2015) arXiv:1412.6980.
\bibitem{awasthi2026pinn} A. Awasthi, Constructing inverse potentials from scattering phase shifts using physics-informed neural networks, 
https://github.com/AayushiAwasthi/Constructing-Inverse-Potentials-from-Scattering-Phase-Shifts-using-Physics-Informed-Neural-Networks
(2026).
\bibitem{adler2017solving}J. Adler and O. ¨Oktem, Solving ill-posed inverse problems using iterative deep neural networks, Inverse Problems 33 (2017).
\bibitem{Ajzenberg1979}F. Ajzenberg-Selove, Energy levels of light nuclei A = 5–10, Nuclear Physics A 320, 1 (1979).
\bibitem{Borsaru1977}M. Borsaru and R. E. Meads, R-function analysis of
$^{4}He + n$ scattering below 21 MeV, Nuclear Physics A 292,
61 (1977).
\bibitem{Barker1985}F. C. Barker and A. C. L. Barnard, Channel radius dependence in R-matrix analysis of light nuclei, Nuclear Physics A 440, 269 (1985).
\bibitem{Arndt1973}R. A. Arndt, D. D. Long, and L. D. Roper, Nucleon-
alpha elastic scattering analyses:(i). low-energy $n-\alpha$ and $p-\alpha$ analyses, Nuclear Physics A 209, 429 (1973).
\bibitem{Tarantola2005}A. Tarantola, Inverse Problem Theory and Methods for Model Parameter Estimation (SIAM, Philadelphia, 2005).

\end{thebibliography}
\end{document}